\documentclass[oneside,12pt,reqno]{article}

\usepackage{latexsym}
\usepackage{amsmath}
\usepackage{amssymb}
\usepackage{amsfonts}
\usepackage{multirow}
\usepackage{geometry}

\iffalse

\else

\fi

\newtheorem{thm}{Theorem}[section]
\newtheorem{lem}[thm]{Lemma}
\newtheorem{cor}[thm]{Corollary}
\newtheorem{prop}[thm]{Proposition}

\newtheorem{dfn}[thm]{Defintion}

\numberwithin{equation}{section}

\newcommand{\mx}{{Hierarchical  }}

\newcommand{\nn}{\nonumber}
\newcommand{\ds}{\displaystyle}
\newcommand{\tab}{\hspace{.025\textwidth}}

\begin{document}

\iffalse

\begin{center}
\Large {\bf Asset Pricing with Random Volatility}\\
 \tab\\
\large {Xin Liu\footnote[1]{Independent Researcher. Email: liuxinrn@gmail.com
}}\\
 \tab\\
%First Version: July 15, 2016; Revised Version: \today
\end{center}

\baselineskip=23pt 

\newpage

\begin{abstract}
This paper proposes to model asset price dynamics with a mixture of diffusion processes where the instantaneous volatility of the underlying diffusion process contains a random vector. The marginal probability distributions of the proposed process can match exactly the risk-neutral distributions implied by both spot vanilla options and forward start options. We can also derive the explicit pricing formula for derivatives that have a closed-form solution under Generalized Geometric Brownian Motion.  \\


$$ $$

\noindent{\it JEL Index:} G12, G13


\end{abstract}

\else

\begin{center}
\Large {\bf Asset Pricing with Random Volatility}\\
 \tab\\
\large {Xin Liu\footnote[1]{Independent Researcher. Email: liuxinrn@gmail.com
}}\\
 \tab\\
First Version: July 15, 2016; Revised Version: \today
\end{center}

$$\\ \\ $$

\baselineskip=23pt

\begin{abstract}
This paper proposes to model asset price dynamics with a mixture of diffusion processes where the instantaneous volatility of the underlying diffusion process contains a random vector. The marginal probability distributions of the proposed process can match exactly the risk-neutral distributions implied by both spot vanilla options and forward start options. We can also derive the explicit pricing formula for derivatives that have a closed-form solution under Generalized Geometric Brownian Motion.  \\


\end{abstract}

\thispagestyle{empty}

\fi

\thispagestyle{empty}

\newpage

\section{Introduction}

It is challenging to calibrate the commonly used asset pricing model to fit both the forward implied volatility and the spot implied volatility. Local volatility models can be formulated to match the spot volatility exactly (\cite{Dupire1994}). However, it does not price forward start options correctly because it has wrong forward dynamics (\cite{Hagan2002}, \cite{Gatheral2006}). As pointed by \cite{Bergomi2005}, the commonly used stochastic volatility models have intractable forward implied volatility. It is thus difficult to obtain the suitable parametrization for stochastic volatility models that can capture the dynamics of the forward start options. \cite{Cont2013} shows that these stochastic models are not capable of reproducing the skew implied by VIX Option. \cite{Behvand2010} demonstrates that the time-dependent Heston stochastic volatility model (\cite{Heston1993}) can provide good fits for either the spot vanilla options or forward start options, but not both. To overcome the constraints of traditional stochastic volatility and jump diffusion models in fitting the Vol-of-Vol structures, \cite{Bergomi2005}, \cite{Bergomi2008} and \cite{Cont2013} model the variance of asset prices with multi-factor term structure models. 

This paper adapts the mixture model approach to model asset price dynamics. Mixture model is an effective tool for analyzing financial time series data because it is flexible enough to capture the idiosyncrasies of financial data (\cite{McLachlan2004}). Particularly, the finite mixture models provide an exact solution for the asset price dynamics with intuitively appealing features by using weighted sums of Black-Scholes solutions. \cite{Ritchey1990} suggests that the observed fat-tailed and skewed distributions can be modeled with finite normal mixture of independent Gaussian processes. \cite{Melick1997} obtains similar findings for the American-style options on crude oil futures. \cite{Alexander2004} introduces a parametrization of the normal mixture diffusion model that captures the short-term and long-term smile effect. \cite{Jacquier1994} proposes to estimate the mixing function using Bayesian analysis. \cite{Gulisashvili2012} obtains the formula of mixing distributions for various diffusions with correlated or uncorrelated stochastic volatilities. \cite{Brigo2002} and \cite{Brigo20022} derive a stochastic differential equation (SDE) whose density evolves as a finite mixture of Gaussian densities.

This paper shows how to derive the stochastic process with an infinite number of mixture components, called mixture diffusion, such that its marginal distribution function fits precisely the risk-neutral distributions implied by spot vanilla options and forward start options. The mixture diffusion is a stochastic process where the instantaneous drift and volatility of the underlying diffusion process contain a random vector. We call this modeling method Random Volatility Models.

The mixture diffusion are shown to be mathematically workable and we can explicitly solve the SDE for mixture diffusion with random volatility. When the dimension of the random volatilities is one, the calibrated mixture diffusion can have the same marginal distribution as the risk-neutral distribution implied by spot vanilla options. The mixture diffusion shares the properties similar to those of local volatility models (\cite{Dupire1994}). Unfortunately, it also inherits its limitations such as flawed forward dynamics (\cite{Hagan2002}, \cite{Gatheral2006}). To overcome these limitations, we propose to model the mixture diffusion with multi-dimensional random volatility called \mx Mixture Diffusion. \mx Mixture Diffusion can precisely model the Vol-of-Vol distribution derived from forward start options and we are able to have an exact fit to the risk-neutral distributions implied by both spot vanilla options and forward start options.

Random Volatility Model is a very general approach in modeling asset dynamics. Many of well-known asset pricing models can be derived directly from Random Volatility Models. For example, Lognormal-Mixture Model in \cite{Brigo20022} and Local Volatility Model in \cite{Dupire1994} can be derived as the Markovian Projection of Random Volatility Models. The posterior distribution obtained via Bayesian analysis in \cite{Jacquier1994} is a special case of Random Volatility Model where the mixing function is parametrized with a finite number of parameters.

Furthermore, Random Volatility Model not only retains the strength of these tradition models but also eschews flaws in their model dynamics. As an example, we consider the mixture diffusion that yields Dupire's Local Volatility Model with Markovian Projection. The discussion following Proposition \ref{propthm3x} shows that its implied volatility smile/skew moves in parallel with the price of the underlying asset. On the other hand, Dupire's Local Volatility Models moves in opposite to typical market behavior (\cite{Hagan2002}). With \mx Random Volatility Model, we can precisely model the Vol-of-Vol Structure of forward dynamics while Local Volatility Model has incorrect forward dynamics.

Below is a short list of the contributions of the proposed modeling method.

\begin{itemize}
\item[1] Under mixture diffusion, the marginal probability distributions of the asset price can match exactly the risk-neutral distributions implied by both spot vanilla options and forward start options, across all expiries and maturities.
\item[2] We have explicit formula for the SDE of mixture diffusion that is consistent with the given risk-neutral distributions. We can also derive the explicit pricing formula for derivatives that have a closed-form solution under Generalized Geometric Brownian Motion.
\item[3] Mixture diffusion can have the same finite joint probability distribution function as that of stochastic volatility models at a fixed set of maturities. Consequence, the mixture diffusion can price derivatives identically as stochastic volatility models if the values of derivatives are completely determined at those given times.
\end{itemize}

\section{Random Volatility Model} \label{sec22}

\subsection{Definition of Mixture Diffusion} \label{sec30}

We explore various methods to find stochastic process in the probability space $(\Omega, \mathcal{F},\{\mathcal{F}_t\}_{t\ge t_0},\mathbb{Q})$ such that its marginal distribution functions match the risk-neutral distributions, where the initial time $t_0\ge0$ is typically $0$. Because we concern only with fitting the risk-neutral distribution, we assume $\mathbb{Q}$ is the same risk-neutral measure from which the risk-neutral probability density functions are derived. We assume the numeraire corresponding to the risk-neutral measure is the money market account $B(t)$, where $B(t):=e^{\int_{t_0}^t r(t)\,ds}$ and $r(t)$ is the risk-free interest rate. We will derive the risk-neutral asset dynamics directly under $\mathbb{Q}$ and our approach does not involve the physical measure.

Let $\Theta$ be a Lebesgue measurable set in $\mathbb{R}^n$ for some $n>0$ and $\tau_0>0$ be a fixed future time. For every $\theta\in\Theta$, we assume that $\mu(x,t; \theta)$ and $\nu(x,t; \theta)$ are scalar functions such that the unique strong solution exists for the SDE
\begin{eqnarray} 
\begin{array}{rcl}
	\,dX_t(\theta)&=&\mu(X_t(\theta),t; \theta)\,dt+\sqrt{\nu(X_t(\theta),t; \theta)}\,dW_t\\
	X_{t_0}(\theta)&=&x_0
\end{array} \label{SDE}
\end{eqnarray}
where $t\in[t_0, \tau_0]$. We denote the support interval of $X_t(\theta)$ as $(a_0,\infty)$ where $a_0$ is typically either $0$ or $-\infty$. We also assume the existence of the marginal probability density function of $X_t(\theta)$ with respect to the Lebesgue measure in $(a_0,\infty)$. We denote $P(\cdot,t;\theta)$ as the density function at time $t$ with the parameter $\theta$.

In the case of the finite mixture model, $\Theta$ is discrete, i.e., $\Theta=\{\theta_1,\ldots,\theta_k\}$. The mixture distribution based on $X_t(\theta)$ is defined as
$$	\tilde{P}(x,t)=\sum_{i=1}^k\lambda_iP(x,t;\theta_i),$$ 
where the mixture proportions $\{\lambda_1,\ldots,\lambda_k\}$ satisfies $\lambda_i\ge0$ and $\sum_{i=1}^k\lambda_i=1$. Equivalently, we can write the mixture proportion in a functional form, denoted as $m(\theta)$,
\begin{eqnarray}
	m(\theta)=\sum_{i=1}^k\lambda_i\delta(\theta-\theta_i)\nn
\end{eqnarray}
where $\delta(\cdot)$ is the Dirac Delta function. Then the mixture distribution becomes,
\begin{eqnarray} 
	\tilde{P}(x,t)=\int_{\Theta}m(\theta)P(x,t;\theta)\,d\theta\nn.
\end{eqnarray} 
In this paper, we will use the functional form of mixture proportion as it is adaptive to infinite number of mixture components.

\begin{dfn}\label{mixtureDF} Assume $(\Omega, \Sigma, m)$ is a Lebesgue measure space. We call 
\begin{eqnarray} 
	\tilde{P}(x,t)=\int_{\Theta}P(x,t;\theta)m(d\theta)\label{mixdist}.
\end{eqnarray} 
the {\it Mixture Distribution} of the parametrized stochastic processes $X_t(\theta)$, and 
call $m$ the {\it Mixing Function} for the parametrized diffusion process \eqref{SDE}.  We call a stochastic process $\{\tilde{X}_t: t_0\le t\le \tau_0\}$ the {\it Mixture Diffusion} of $\tilde{P}(x,t)$ if its marginal probability density function is the mixture distribution $\tilde{P}(\cdot,t)$ for every $t_0< t\le \tau_0$. 
\end{dfn}

It is clear that our definition of mixing function ensures that $\tilde{P}(x,t)$ is a density function with the support interval $(a_0,\infty)$. \cite{Brigo2002} derives the explicit formula for finite mixture diffusions when the drift and volatility terms are deterministic functions of the underlying asset price. This paper proposes to model asset price with the mixture of infinite diffusions where the drift and volatility terms contain random vectors. Let $(\Omega^w, \mathcal{F}^w,\{\mathcal{F}^w_t\}_{t\ge0}, \mathbb{P}^w)$ be the filtered probability space for the standard Brownian motion $\{W_t, t\ge0\}$; $V$ be a $n$-dimensional random vector with the probability distribution $m(\cdot)$ on the probability space $(\Omega^v, \mathcal{F}^v, \mathbb{P}^v)$. We assume $V$ is independent to $W_t$. Let $\mathbb{Q}$ be the product measure for $\mathcal{F}^w\otimes\mathcal{F}^v$. Denote the probability space with product measure $\mathbb{Q}$ as $(\Omega, \mathcal{F},\{\mathcal{F}_t\}_{t\ge t_0}, \mathbb{Q}):=(\Omega^w\times\Omega^v, \mathcal{F}^w\otimes\mathcal{F}^v,\{\mathcal{F}^w_t\otimes\mathcal{F}^v\}_{t\ge t_0}, \mathbb{Q})$. On the probability space $(\Omega, \mathcal{F},\{\mathcal{F}_t\}_{t\ge t_0}, \mathbb{Q})$, we define the mixture diffusion with random drift and volatility based on the parametrization \eqref{SDE} as
\begin{eqnarray}
\begin{array}{rcl}
	d\tilde{X}_t&=&\mu(\tilde{X}_t,t;V)\,dt+\sqrt{\nu(\tilde{X}_t,t;V)}\,dW_t\\
	\tilde{X}_{t_0}&=&x_0
\end{array}	\label{SDE_R4}
\end{eqnarray}
Without loss of generality, we assume that $X_t(\theta)$ resides in the same probability space of $\tilde{X}_t$ such that $\tilde{X_t}=X_t(\theta)$ when $V=\theta$, i.e.,
\begin{eqnarray}
	\tilde{X}_t=X_{t}(V).\label{SDE_R5}
\end{eqnarray}
Then the marginal distribution function of $\tilde{X}_t$ is,
\begin{eqnarray}
\mathbb{P}\big(\tilde{X}_t\in\,dx\big)
%&=&\int_\Theta\mathbb{P}\big(\tilde{X}_t\in\,dx|V=\theta\big)\mathbb{P}\big(V\in\,d\theta\big)\nn\\
=\int_\Theta\mathbb{P}\big({X}_t(V)\in\,dx|V=\theta\big)\mathbb{P}\big(V\in\,d\theta\big)
=\tilde{P}(x,t)\,dx\nn.
\end{eqnarray}
Therefore $\tilde{X}_t$ satisfies the definition of mixture diffusion for $\tilde{P}(x,t)$.

The mixture diffusion defined in \eqref{SDE_R4} is in fact a Hidden Markov Process, in which $\tilde{X}_t$ represents the observed asset price and $V$ represents the unobserved state variable. The hidden state variable can be interpreted as a random sample from the universe of investors. Each individual investor has his/her own idiosyncratic view on the asset evolution and this is modeled with the parametrized stochastic process \eqref{SDE}. However, what is observable to market participants is the aggregated market movement as the results of the actions of all investors. The aggregated asset price movements thus follow the hidden Markov process with the hidden state variable sampled from the population of investors. Consequently, under mixture diffusions, the asset price itself is not a Markov process.

In the paper, we call the mixture diffusion with the instantaneous drift term $r(t)x$ as risk-neutral mixture diffusion. Under the risk-neutral mixture diffusion, we consider the price of the derivative with the payoff function $C(\cdot):S^T\mapsto\mathbb{R}$, where $S=(0,\infty)$ is the state space of $\tilde X_t$ and $T=[t_0,\tau_0]$ is its index set. Note that $\tilde{X}_t$ and $X_t(\theta)$ can be regarded as $S^T$--valued random variables. The expected value of derivative becomes
\begin{eqnarray} 
\mathbb{E}[C(\tilde{X})]
=\mathbb{E}(\mathbb{E}[C(\tilde{X})|V])
=\int_\Theta \mathbb{E}[C({X}(\theta))]m(d\theta).\label{CCValue}
\end{eqnarray}
Equation \eqref{CCValue} gives the important property in the valuation of derivatives: the value of a derivative under mixture diffusion equals the weighted average of the expected values across all underlying diffusions. With \eqref{CCValue}, we can obtain the closed-form solution of exotic options if they have explicit solution for the underlying diffusions. Another implication of \eqref{CCValue} is that Greeks (or hedging parameters) can be directly calculated as the weighted average of Greeks across all underlying diffusions, i.e.,
\begin{eqnarray} 
\frac{\partial}{\partial x}\mathbb{E}[C(\tilde{X})]
=\int_\Theta \frac{\partial}{\partial x}\mathbb{E}[C({X}(\theta))]m(d\theta),\label{CCValue2}
\end{eqnarray}
where $x$ represent an underlying parameter on which the value of the derivative is dependent.

In this paper, we consider one particular risk-neutral parametrization that can be explicitly solved from risk-neutral distributions: the Generalized Geometric Brownian Motion (GGBM)
\begin{eqnarray} 
\begin{array}{lll}
	\ds\frac{\,dX_t(\theta)}{X_t(\theta)}&=&r(t)\,dt+\sqrt{\nu(\theta,t)}\,dW_t\\
	X_{t_0}(\theta)&=&x_0
\end{array}
\label{SDE_GD}
\end{eqnarray}
where $t\in[t_0, \tau_0]$,  $\nu(\theta,t)$ is a deterministic scalar function of $(\theta,t)$.

\begin{dfn}\label{localMD}
For any $t\in(t_0,\tau_0]$, we assume $(\Omega,\Sigma, m_t)$ is a Lebesgue measure space. We call the underlying parametrization of \eqref{SDE_R40} the {\it Mixture of Geometric Parametrization} (MGP). We define the mixture distribution for MGP at time $t$ with the mixing function $m_t$ as $\int_\Theta P(x,t;\theta)m_t(d\theta)$. For simplicity, we express the parametrization as the vector $(m_t,\nu(\theta,t),t_0, x_0, \Theta).$
\end{dfn}

The mixing function in MGP is allowed to be time-dependent because MGP is a tool to model mixture distribution.  Through-out the paper, we apply only time-independent mixing function to the definition of mixture diffusions as below.

\begin{dfn}\label{localMGD}
Let $\mathcal{M}=(m,\nu(\theta,t),t_0, x_0, \Theta)$ and define the mixture diffusion
\begin{eqnarray}
\begin{array}{lcl}
	\ds\frac{\,d\tilde{X}_t}{\tilde{X}_t}&=&r(t)\,dt+\sqrt{\nu(V,t)}\,dW_t\\
	\tilde{X}_{t_0}&=&x_0
\end{array}	\label{SDE_R40}
\end{eqnarray}
where $V$ is the random variable with the probability distribution $m(\cdot)$; we assume $V$ is adapted to $\mathcal{F}_{t_0}$ and is independent of $\{W_t,0\le t\le \tau_0\}$. We call $\tilde{X}_t$ the  {\it Mixture of Geometric Diffusion} (MGD) with the parametrization $\mathcal{M}$. For simplicity, we express the MGD as the vector $(\tilde X_t,\mathcal{M},V)$. %The initial time $t_0$ is $0$ if our focus is spot volatility only.
\end{dfn}

\subsection{Properties of Mixture Diffusion} \label{sec301}

First we derive the sufficient conditions such that the unique strong solution exists for the mixture diffusion \eqref{SDE_R4}. Typically the linear growth and Lipschitz conditions can ensure the existence of the unique strong solution. Though these conditions are usually satisfies by the underlying parametrized diffusion, they fail for the mixture diffusion when the parametrization has unbounded parameters. The theorem below shows that the conditions similar to the classic result still apply to mixture diffusion. Though our result is stated in one-dimensional form, the same result also holds true when $\tilde{X}_t$ and $W_t$ are vectors if we interpret $|\cdot|$ as the Euclidean norm. The proof of the result can be found in Appendix.

\begin{thm} \label{thmR3} 
Let $\sigma(x,t;\theta)=\sqrt{\nu(x,t;\theta)}$ and $t_0=0$. We assume
\begin{itemize}
\item[C1.] There exists a measurable function $f:\Theta\mapsto[0,\infty)$ such that 
\begin{eqnarray}
&&|\mu(x,t;\theta)|+|\sigma(x,t;\theta)|\le f(\theta)(1+|x|)\nn\\
&&|\mu(x_1,t;\theta)-\mu(x_2,t;\theta)|+|\sigma(x_1,t;\theta)-\sigma(x_2,t;\theta)|\le f(\theta)|x_1-x_2|\nn
\end{eqnarray}
\item[C2.] Assume the mixing function $m(\cdot)$ satisfies 
\begin{eqnarray}
\ds\int_\Theta e^{C_0f^2(\theta)} m(d\theta)<\infty\label{thmR3_EQ1}
\end{eqnarray}
where $\ds C_0=10(\tau_0^2+\tau_0)$.
\item[C3.] Let $V$ be a $n$-dimensional random vector with the probability distribution function $m(\cdot)$. We assume $V$ is adapted to $\mathcal{F}_0$ and is independent of $W_t$. 
\end{itemize}
Then the stochastic process
\begin{eqnarray}
\begin{array}{rcl}
	d\tilde{X}_t&=&\mu(\tilde{X}_t,t;V)\,dt+\sqrt{\nu(\tilde{X}_t,t;V)}\,dW_t\\
	\tilde{X}_{0}&=&x_0
\end{array}	\label{SDE_R14}
\end{eqnarray}
admits a unique strong solution with the marginal distribution function $\tilde{P}(x,t)$ \eqref{mixdist}.
\end{thm}

Using Markovian Projection technique (\cite{Gyongy1986}, \cite{Piterbarg2006}), we can derive a diffusion process with deterministic drift and volatility that has the marginal probability distribution  $\tilde{P}(x,t)$. Below we directly prove this result by verifying the Fokker--Planck equation as \cite{Brigo2002} did.
\begin{prop} \label{propR3} 
We assume the stochastic process $\tilde{X}_t$ satisfies Theorem \ref{thmR3}. Let 
\begin{eqnarray}
\hat{\mu}(x,t)&=& \mathbb{E}(\mu(\tilde{X}_t,t;V)|\tilde{X}_t=x)=\frac{1}{\tilde{P}(x,t)}\int_\Theta \mu(\tilde{X}_t,t;\theta)P(x,t;\theta)m(d\theta),\nn\\
\hat{\nu}(x,t)&=&\mathbb{E}(\nu(\tilde{X}_t,t;V)|\tilde{X}_t=x)=\frac{1}{\tilde{P}(x,t)}\int_\Theta \nu(\tilde{X}_t,t;\theta)P(x,t;\theta)m(d\theta).\nn
\end{eqnarray}
Then the diffusion process $\hat{X}_t$ has the same marginal probability distribution function $\tilde{P}(x,t)$, where
\begin{eqnarray}
\begin{array}{rcl}
	d\hat{X}_t&=&\hat\mu(\hat{X}_t,t)\,dt+\sqrt{\hat\nu(\hat{X}_t,t)}\,dW_t\\
	\hat{X}_{t_0}&=&x_0
\end{array}	\label{SDE_T14}
\end{eqnarray}
\end{prop}

{\noindent\bf Proof of Proposition \ref{propR3}:} We assume that the exchanges of derivation and integral are all valid in the following equations. 
\begin{eqnarray}
	&&\frac{\partial }{\partial t}\tilde{P}(x,t)\nn\\
	&=&\int_\Theta\frac{\partial }{\partial t}P(x,t;\theta)m(d\theta)\nn\\
	&=&\int_\Theta\left(-\frac{\partial}{\partial x}\Big({\mu}(x,t;\theta){P}(x,t;\theta)\Big)
	+\frac12\frac{\partial^2}{\partial x^2}\Big({\nu}(x,t;\theta){P}(x,t;\theta)\Big)\right)m(d\theta)\nn\\
	&=&-\frac{\partial}{\partial x}\int_\Theta {\mu}(x,t;\theta){P}(x,t;\theta)m(d\theta)
	+\frac12\frac{\partial^2}{\partial x^2}\int_\Theta m(\theta){\nu}(x,t;\theta){P}(x,t;\theta)m(d\theta)\nn\\
	&=&-\frac{\partial}{\partial x}\Big(\mathbb{E}(\mu(\tilde{X}_t,t;V)|\tilde{X}_t=x)\tilde{P}(x,t)\Big)
	+\frac12\frac{\partial^2}{\partial x^2}\Big(\mathbb{E}(\nu(\tilde{X}_t,t;V)|\tilde{X}_t=x)\tilde{P}(x,t)\Big)\nn
\end{eqnarray}
Then Fokker-Planck equation ensures that $\tilde{P}(x,t)$ is the density function of $\hat{X}_t$. This completes our proof. \noindent$\blacksquare$

It's well-known that there are infinite solutions of \eqref{SDE_T14} that can have the same marginal probability distribution function $\tilde{P}(x,t)$. We choose the particular form of $\hat{\mu}(x,t)$ and $\hat{\nu}(x,t)$ defined in Proposition \ref{propR3} because we want $\hat{X}_t$ to be risk-neutral for our modeling choice MGD. We also want to point out that the lognormal-mixture dynamics presented in Proposition 3.1 of \cite{Brigo20022} is a special case of Markovian Projection of mixture diffusion. In fact, we can set the mixing function and volatility function of MGD \eqref{SDE_GD} the same as that in \cite{Brigo20022}. Then the direct comparison can show that Markovian Projection \eqref{SDE_T14} is identical to the diffusion process in Proposition 3.1 of \cite{Brigo20022}.

Because $\tilde{X}_t$ is not Markovian, $\tilde{X}_t$ conditioning on $\tilde{X}_s$ is no longer the same stochastic process. However, the constrained mixture diffusion is still a mixture diffusion. More specifically, we can show that the conditional law of mixture diffusion is identical to the law of another mixture diffusion with the same parametrization but a different mixing function. Denote $S=(0,\infty)$ as the state space of $\tilde X_t$; let $I_a=[t_0,t_1)$ and $I_b=(t_1,\tau_0]$ be two index sets; $\mathcal{F}^{a}$ and $\mathcal{F}^{b}$ be the cylinder $\sigma$--algebra of $S^{I_a}$ and $S^{I_b}$, respectively. 
\begin{prop} \label{propR4} Assume $\mathbb{P}(\{\tilde{X}\in A\}\cap\{\tilde{X}_{t_1}=x_1\})>0$ for some  $A\in \mathcal{F}^a$. Then for any $B\in\mathcal{F}^b$, we have, 
$$\mathbb{P}(\tilde{X}\in B|\tilde{X}\in A,\tilde{X}_{t_1}=x_1)=\mathbb{P}(\hat{X}\in B)$$ 
where the mixture diffusion $\{\hat{X}_t,t_1\le t\le\tau_0\}$ is defined as
\begin{eqnarray}
\begin{array}{rcl}
	d\hat{X}_t&=&\mu(\hat{X}_t,t;\hat V)\,dt+\sqrt{\nu(\hat{X}_t,t;\hat V)}\,dW_t\\
	\hat{X}_{t_1}&=&x_1
\end{array}	\label{SDE_R15}\nn
\end{eqnarray}
and the probability distribution of $\hat V$ is the mixing function $\ds\hat m(\cdot)$, where
$$\ds\hat m(d\theta)=\frac{m(d\theta)\mathbb{P}(\{{X}(\theta)\in A\}\cap\{{X}_{t_1}(\theta)=x_1\})}{\mathbb{P}(\{\tilde{X}\in A\}\cap\{\tilde{X}_{t_1}=x_1\})}.$$
\end{prop}
{\noindent\bf Proof of Proposition \ref{propR4}:} Note that the underlying diffusion $X_t(\theta)$ is Markovian.
\begin{eqnarray}
	&&\mathbb{P}(\tilde{X}\in B|\tilde{X}\in A,\tilde{X}_{t_1}=x_1)\nn\\
	&=&\int_\Theta  \mathbb{P}(\tilde{X}\in B|\tilde{X}\in A,\tilde{X}_{t_1}=x_1,V=\theta)P(V\in\,d\theta)\nn\\
	&=&\int_\Theta  \mathbb{P}({X}(\theta)\in B|{X}(\theta)\in A,{X}_{t_1}(\theta)=x_1)m(d\theta)\nn\\
	&=&\int_\Theta  \mathbb{P}({X}(\theta)\in B|{X}_{t_1}(\theta)=x_1)\hat m(d\theta)\nn\\
	&=&\int_\Theta  \mathbb{P}(\hat{X}\in B|\hat{V}=\theta)P(\hat{V}\in\,d\theta)\nn\\
	&=&\mathbb{P}(\hat{X}\in B)\nn
\end{eqnarray} \noindent$\blacksquare$

The posterior distribution derived by the Bayesian method in \cite{Jacquier1994} is a special case of the mixing function $\hat m(\cdot)$ with $A$ represents the collection of events that yields the observed asset prices. However, we want to point out that the Bayesian method is not the correct way to draw inference on the mixing function. An heuristically explanation is that the true mixing function does not have a parametric functional form. Therefore, the Bayesian method does not apply because it relies on the parameter estimation. Alternatively, we can look at the asymptotic distribution of the posterior distribution. In general, as the number of observations goes to infinity, the posterior distribution in Bayesian analysis typically converges to a Dirac Delta function that centered at certain parameters for the underlying diffusions. Therefore, the Bayesian method is asymptotically equivalent to an optimized estimation of the underlying diffusions and this is distinct from the true mixing diffusion.

\subsection{Explicit Formula of MGD} \label{sec31}

In this section, we focus on deriving the suitable MGD such that its marginal distribution functions fit exactly to risk-neutral distribution across all maturities. Our modeling approach follows these steps: first we choose a generic parametrization based on GGBM. Next we solve for its mixing function at maturity $T$, such that the resulting mixture distribution equals the given risk-neutral distribution. After repeating the second step for every known $T\in(t_0,\tau_0]$, we obtain a time-dependent mixing function for this generic parametrization. Lastly, we will re-parametrize the MGP and convert the mixing function into a time-independent one and obtain the desired MGD as defined in \eqref{SDE_R40}. This re-parametrization is achievable with the next proposition, where we can specify an arbitrary time-independent mixing function, then re-parametrize the previously derived time-dependent MGP in such way that the re-parametrized MGP still have the same marginal distribution. Note that the target mixing function in the result below can be a time-dependent function as well.

\begin{prop}\label{corthm3} For any MGP $\mathcal{M}=(m_t,\nu(\theta,t),t_0,x_0,\Theta)$, we define the re-parametrized MGP as $\hat{\mathcal{M}}=(\hat m_t,\hat\nu(\hat\theta,t),t_0,x_0,\hat\Theta)$, where 
\begin{eqnarray}
	\ds\hat\nu(\hat\theta,t)=\ds\frac{\partial }{\partial t}\int_{t_0}^t\nu(M_t^{-1}(\hat M_t(\hat\theta)),s)\,ds\label{MIXEQU20}
\end{eqnarray} 
and $\hat\nu(\hat\theta,t)$ is assumed to be well-defined. $M_t(\cdot)$ and $\hat M_t(\cdot)$ are the cumulative distribution functions of $m_t(\cdot)$ and $\hat m_t(\cdot)$, respectively; and we assume their inverse functions exist.

Then $\mathcal{M}$ and $\hat{\mathcal{M}}$ have the same mixture distribution for every $t\in(t_0,\tau_0]$. 
\end{prop}

{\noindent\bf Proof of Proposition \ref{corthm3}:} Let $g_t(\theta)=M_t^{-1}(\hat M_t(\theta))$. Define 
$$\ds v(\theta,t)=\ds\int_{t_0}^t\nu(\theta,s)\,ds\tab\textup{and}\tab\ds\hat v(\theta,t)=\ds\int_{t_0}^t\hat \nu(\theta,s)\,ds;$$ 
It is clear that $\hat v(\theta,t)=v(g_t(\theta),t)$ and $\hat u(\theta,t)=u(g_t(\theta),t)$. 
Let $F(t)=x_0\exp(\int_{t_0}^tr(s)\,ds)$ be the forward asset price. Then the mixture distribution of $\mathcal{M}$ is
\begin{eqnarray} 
	&&\int_\Theta \frac{1}{x\sqrt{2\pi v(\theta,t)}}\exp\left(-\frac{(\log(x/F(t))+v(\theta,t)/2)^2}{2v(\theta,t)}\right)\,dM_t(\theta)\nn\\
	&=&\int_{\hat\Theta} \frac{1}{x\sqrt{2\pi v(g_t(\theta),t)}}\exp\left(-\frac{(\log(x/F(t))+v(g_t(\theta),t)/2)^2}{2v(g_t(\theta),t)}\right)\,dM_t(g_t(\theta))\nn\\
	&=&\int_{\hat\Theta}\frac{1}{x\sqrt{2\pi\hat{v}(\theta,t)}}\exp\left(-\frac{(\log(x/F(t))+\hat{v}(\theta,t)/2)^2}{2\hat{v}(\theta,t)}\right)\hat m_t(d\theta)\nn
\end{eqnarray}
This completes our proof. \noindent$\blacksquare$

We call two MGPs equivalent if their parametrization satisfies the relation defined in Proposition \ref{corthm3}. 
\begin{dfn}\label{localSIM}
We call $\mathcal{M}=(m_t,\nu(\theta,t),t_0,x_0,\Theta)$ equivalent to $\mathcal{\hat M}=(\hat m_t,\hat\nu(\hat\theta,t),t_0,x_0,\hat\Theta)$ if the following equation holds
\begin{eqnarray}	
	\ds\int_{t_0}^t\hat\nu(\hat M_t^{-1}(x),s)\,ds=\ds\int_{t_0}^t\nu(M_t^{-1}(x),s)\,ds
\label{MIXEQU21}
\end{eqnarray}
for every $x\in(0,1)$ and $t\in(t_0,\tau_0]$. We denote the equivalence relation as $\mathcal{ M}\sim\mathcal{\hat M}$. 
\end{dfn}

It is straightforward to verify that the equivalence relation $\sim$ defines an unique mixture diffusion in the weaker sense. This also implies that the price of derviatives is invariant under the equivalence relation.  Denote $S=(0,\infty)$ as the state space of $\tilde X_t$, $I_{0}=[t_0,\tau_0]$ as the index set and $\mathcal{F}^{0}$ as the cylinder $\sigma$--algebra of $S^{I_0}$. 
\begin{prop}\label{corthm4}
Consider two MGD $(\tilde X_t,\mathcal{M},V)$ and $(\hat X_t,\hat{\mathcal{M}},\hat V)$ with $\mathcal{M}\sim\hat{\mathcal{M}}$. Then for any $A\in\mathcal{F}^{0}$, we have  $\mathbb{P}(\tilde{X}\in A)=\mathbb{P}(\hat X\in A).$   
\end{prop}
{\noindent\bf Proof of Proposition \ref{corthm4}:} We denote the underlying parametrized diffusion process of $\mathcal{M}$ and $\hat{\mathcal{M}}$ as $X^1_t(\theta)$ and $X^2_t(\hat\theta)$, respectively. Then \eqref{MIXEQU21} implies that, $X^1_t(M^{-1}(x))$ and $X^2_t(\hat M^{-1}(x))$ has the same law for every $x\in(0,1)$. Consequently, we have
\begin{eqnarray}	
	\mathbb{P}(\tilde{X}\in A)
	&=&\int_\Theta m(\theta)\mathbb{P}(X^1(\theta)\in A)\,d\theta
	=\int_0^1 \mathbb{P}(X^1(M^{-1}(x))\in A)\,dx\nn\\
	&=&\int_0^1 \mathbb{P}(X^2(\hat M^{-1}(x))\in A)\,dx
	=\mathbb{P}(\hat{X}\in A)\nn
\end{eqnarray} 
\noindent$\blacksquare$

Finally we can present the explicit formula of MGD whose mixture distributions equal given riks-neutral distributions. To do so, first we derive the parametrization
$\mathcal{M}=(m_{t},\theta,t_0, x_0,\mathbb{R}^+)$ such that the mixture distribution of $\mathcal{M}$ equals the given risk-neutral distribution. Note that the mixture distribution of $\mathcal{M}$ at maturity $t$ is 
$$\tilde{P}(x,t)=\int_0^\infty \frac{m_t(\theta)}{\sqrt{2\pi\theta t}x}\exp\left(\frac{(\log(x/F(t))-\theta t/2)^2}{2\theta t}\right)\,d\theta,$$
where $F(t)=x_0\exp(\int_{t_0}^tr(s)\,ds)$ is the forward asset price. It is more convenient to consider the marginal distribution based on the logarithm of moneyness $y:=\log(x/F(t))$. We denote it as $\tilde{Q}(y,t)$:
$$\tilde{Q}(y,t):=\int_0^\infty \frac{m_t(\theta)}{\sqrt{2\pi\theta t}}\exp\left(\frac{(y-\theta t/2)^2}{2\theta t}\right)\,d\theta.$$
Denote $D_t(x)$ as the risk-neutral distribution of the asset price and
$E_t(x)$ as the risk-neutral distribution for the logarithm of moneyness of the asset price: $$E_t(x):=F(t)e^xD_t(F(t)e^x).$$

Then $\tilde{P}(x,t)$ equals the risk-neutral distribution $D_t(x)$ if and only if $\tilde{Q}(x,t)$ equals $E_t(x)$. Taking Fourier transfer of both $\tilde{Q}(y,t)$ and $E_t(y)$, it yields the equation
\begin{eqnarray}
\mathcal{F}(E_t)(\eta)=\int_0^\infty m_t(\theta)\exp\left(-(i\eta+\eta^2)\theta t/2\right)\,d\theta.\label{fourierequ1}
\end{eqnarray}
The necessary and sufficient condition for existence of mixing function $m_t(\cdot)$ satisfying \eqref{fourierequ1} is that, $G_t(\cdot)$ is completely monotonic \cite{Widder1941}, where
\begin{eqnarray}
G_t(\eta):=\mathcal{F}(E_t)\left(\sqrt{\frac{2\eta}{t}-\frac14}-\frac{i}2\right).\label{fouriersolution0}
\end{eqnarray} 
Under the sufficient condition, the Laplace inversion of $G_t(\cdot)$ exists and the unique solution of mixing function $m_t(\cdot)$ is 
\begin{eqnarray}
m_t(x)=\mathcal{L}^{-1}(G_t)(x).\label{fouriersolution1}
\end{eqnarray}

Lastly we can determine the desired risk-neutral MGD $(\hat{X}_t,\hat{\mathcal{M}},\hat V)$ by solving $\hat{\mathcal{M}}=(\hat m,\hat\nu(\hat\theta,t),t_0,x_0,\hat\Theta)$ from the equivalent relation $\mathcal{M}\sim\hat{\mathcal{M}}$. Denote the cumulative distribution functions of $m_t(\cdot)$ and $\hat m(\cdot)$ as $M_t(\cdot)$ and $\hat M(\cdot)$, respectively. Based on the notations from Proposition \ref{corthm3}, the explicit solution of $\hat\nu(x,t)$ is $$\hat\nu(x,t)=\frac{\partial}{\partial t}M^{-1}_t(\hat M(x))=\frac{\partial}{\partial t}Z^{-1}_t(\hat M(x)),$$ where $Z_t(x)=\int_0^x\mathcal{L}^{-1}(G_t)(y)\,dy.$

We conclude this section by summarize the derivation above into the following proposition. Without loss of generality, we set the mixing function as the uniform distribution.

\begin{prop}\label{propthm3x} Assume that $G_t(\cdot)$ \eqref{fouriersolution0} is completely monotonic and the volatility term $\hat\nu(x,t)=\frac{\partial}{\partial t}Z^{-1}_t(x)$ is well-defined. Define $\hat{\mathcal{M}}=(1,\hat\nu(x,t),t_0,x_0,[0,1])$. Then MGD $(\hat{X}_t,\hat{\mathcal{M}},\hat V )$ has the marginal distribution function $D_t(\cdot)$ for every $t\in(t_0,\tau_0]$.
\end{prop}

\cite{Dupire1994} shows that there is an unique formula for the volatility function in the risk-neutral Local Volatility Model when its marginal distribution equals the risk-neutral distribution across all maturities. Note that the Markovian Projection of the MGD derived in Proposition \ref{propthm3x} is a risk-neutral Local Volatility Model and its marginal distribution equals the risk neutral distribution $D_t(\cdot)$ for every $t\in(t_0,\tau_0]$. Therefore, the Local Volatility Model derived in \cite{Dupire1994} is identical to the Markovian Projection of MGD in Proposition \ref{propthm3x}. As pointed out by \cite{Hagan2002}, Local Volatility Models predict that the market
smile/skew moves in the opposite direction as the price of the underlying asset, which is opposite to typical market behavior. In contrast, we can show that the Random Volatility model above has the correct Delta risk. Note that GGBM has sticky Delta, i.e., its value of vanilla European options is a function of the moneyness $\ln(K/S_0)$, where $K$ represents the strike and $S_0$ represents the initial asset price. Then \eqref{CCValue} implies that MGD has sticky Delta as well. Therefore, the smile/skew of MGD moves in the parallel with the price of the underlying asset and this is consistent with  market behavior.

\section{\mx Random Volatility Model} \label{sec5}

The mixture diffusion derived in Proposition \ref{propthm3x} has consistent prices for spot vanilla options across all maturities. However, it cannot price forward start option correctly because its forward implied volatility tends to be flat. For example, we can consider the forward implied volatility of this MGD given $\mathcal{F}_s$. Because $\tilde X_s$ are $V$ are known at time $s$, the constrained stochastic process becomes the underlying GGBM with the parameter setting to $V$ and starting value to $\tilde{X}_s$. Consequently, its forward implied volatility equals the spot implied volatility of this GGBM and its volatility surface is thus flat.

In this section, we show how to re-construct mixture diffusion at forward times so that the expected volatility can match both the forward implied volatility and spot implied volatility. At each forward time, re-constructing mixture diffusion is based on the mixture distribution at prior times, just as neutral network models with multiple layers. We call this approach \mx Mixture Diffusion to highlight the layered parametrization approach. Similarly, we call the corresponding parametrization as \mx MGP and this type Random Volatility Model as \mx Random Volatility Model.

\subsection{Definition of \mx Mixture Diffusion} \label{sec51}

Let $T_0,\cdots,T_{n+1}$ be a series of fixed times that satisfy $0\le T_0<T_1<\cdots<T_{n+1}\le\tau_0$. Our \mx MGP is based on a series of MGP defined for the time interval $[T_{k},T_{k+1}]$ conditioning on total variance at $(T_0,\ldots,T_{k})$ for every $k=0,\ldots,n$. We use the symbol $\boldsymbol{\sigma}_k=(\sigma_0,\sigma_{1},\ldots,\sigma_{k})$ denote the sequence of total variance at $(T_0,T_1,\ldots,T_k)$, where $\sigma_0\ge0$ is a constant known at the starting time $T_0$. We assume that the total variance sequence satisfies the constraint $\sigma_0\le\sigma_{1}\le\cdots\le\sigma_{n}$. Then we can write the MGP for the time interval $[T_{k-1},T_{k}]$ as
\begin{eqnarray}
	\mathcal{M}_k=(m_k,\nu_k(\theta_k,t;\boldsymbol{\sigma}_{k-1}),T_{k-1},1,\Theta_k)\label{modelvariance00}
\end{eqnarray}

Note that we do not treat $\boldsymbol{\sigma}_{k-1}$ as free parameters because we can explicitly formulate the total variance as a function of $(\theta_{1},\ldots,\theta_{k-1})$. Let $\boldsymbol{\theta}_k=(\theta_{1},\ldots,\theta_{k})$. We denote $v_{k}(\boldsymbol{\theta}_{k})$ as the sum of variance from the parameterizations $\mathcal{M}_1,\ldots,\mathcal{M}_{k}$, and let $\boldsymbol{v}_{k}(\boldsymbol{\theta}_{k})=(v_0,v_{1}(\boldsymbol{\theta}_1),\ldots,v_{k}(\boldsymbol{\theta}_{k}))$ where we assume the initial variance $v_0\ge0$ is a known constant. Note that the variance contributed by $\mathcal{M}_k$ is  $\int_{T_{k-1}}^{T_{k}}\nu_{k}(\theta_{k},t;\boldsymbol{v}_{k-1}(\boldsymbol{\theta}_{k-1}))\,ds$. Therefore, $v_{k}(\boldsymbol{\theta}_{k})$ satisfies the equation 
\begin{eqnarray}
v_{k}(\boldsymbol{\theta}_{k})=v_{k-1}(\boldsymbol{\theta}_{k-1})+ \int_{T_{k-1}}^{T_{k}}\nu_{k}(\theta_{k},t;\boldsymbol{v}_{k-1}(\boldsymbol{\theta}_{k-1}))\,ds\label{modelvariance1}
\end{eqnarray} 
Below we define \mx MGP based the series of parametrizations in \eqref{modelvariance00} with the total variance defined as in \eqref{modelvariance1}.. 

\begin{dfn}\label{localFWD} We call $\mathcal{M}=(m, \nu(\boldsymbol{\theta},t),T_0,x_0,\Theta)$ the \mx MGP based on the series of parametrization $\{\mathcal{M}_k\}_{k=1}^n$ defined in \eqref{modelvariance00} if
\begin{itemize}
\item[I.] The mixing function satisfies $m(d\boldsymbol{\theta})=\prod_{i=1}^n m_k\left(d\theta_{k}\right)$, where $\boldsymbol{\theta}=(\theta_{1},\ldots,\theta_{n})$ and $\Theta=\Theta_1\otimes\cdots\otimes\Theta_n$;
\item[II.] $v_{k}(\boldsymbol{\theta}_{k})$ satisfies \eqref{modelvariance1} for $k=1,\ldots,n$;
\item[III.] The parametrization satisfies $\nu(\boldsymbol{\theta} ,t)=\nu_k(\theta_{k},t;\boldsymbol{v}_{k-1}(\boldsymbol{\theta}_{k-1}))$, for $t\in[T_{k-1},T_{k})$ and $k=1,\ldots,n$.
\end{itemize}
\end{dfn}

Then \mx MGP can uniquely define a mixture diffusion with random volatility and we call it \mx MGD.
\begin{dfn}\label{mxMGD} 
Assume $\mathcal{M}$ is the \mx MGP defined in Definition \ref{localFWD}; $V_{k}$ is a random variable with probability distribution $m_k(\cdot)$ and is adapted to the filtration $\mathcal{F}_{T_{k-1}}$; $V_{1},\ldots,V_{n},W_t$ are independent of each other. Let $\boldsymbol{V}_k=(V_{1},\ldots,V_{k})$ and $\boldsymbol{V}=\boldsymbol{V}_n$.

Define the stochastic process $\tilde{X}_t$ via SDE
\begin{eqnarray}
\begin{array}{lcl}
	\ds\frac{\,d\tilde{X}_t}{\tilde{X}_t}&=&r(t)\,dt+\sqrt{\nu_k({V}_{k},t;\boldsymbol{v}_{k-1}(\boldsymbol{V}_{k-1}))}\,dW_t\\
	\tilde{X}_{T_0}&=&x_0
\end{array}	\label{SDE_mx40}
\end{eqnarray}
for $t\in[T_{k-1},T_k)$ and $k=1,\ldots,n$. We call $\tilde{X}_t$ the \mx MGD based on the parametrization ${\mathcal{M}}$ and denote SDE \eqref{SDE_mx40} as $(\tilde{X},{\mathcal{M}},\boldsymbol{V})$.
\end{dfn}

For single-layer MGD, the constrained process given $\mathcal{F}_s$ reduces to the underlying GGBM and the surface of the conditional implied volatility is thus flat. In contrast, the constrained \mx MGD has the same probability law as another \mx MGD and the surface of the conditional volatility can be identical to the spot volatility surface. As an example, we consider the conditional probability law of $\tilde{X}_t$ in \eqref{SDE_mx40} with respect to $\mathcal{F}_{T_{k-1}^-}$, where $k \ge 2$. Denote $S=(0,\infty)$ as the state space of $\tilde X_t$; $\mathcal{F}_{k}$ as the cylinder $\sigma$--algebra of $S^{(T_k,T_{n+1}]}$. For any $A\in\mathcal{F}_k$, the conditional probability equals
\begin{eqnarray}
\mathbb{P}\left(\tilde{X}\in A\left|\mathcal{F}_{T_{k-1}^-}\right.\right)
&=&\mathbb{P}(\tilde{X}\in A|\tilde{X}_{T_{k-1}},\boldsymbol{V}_{k-1})\nn\\
&=&\mathbb{P}(X(\boldsymbol{V})\in A|X_{T_{k-1}}(\boldsymbol{V})=\tilde{X}_{T_{k-1}},\boldsymbol{v}_{k-1}(\boldsymbol{V}_{k-1})).\label{falvol2}
\end{eqnarray}
where $X_t(\cdot)$ is the underlying the diffusion process and it satisfies
\begin{eqnarray}
\begin{array}{lcl}
	\ds\frac{\,d{X}_t(\boldsymbol{\theta})}{{X}_t(\boldsymbol{\theta})}&=&r(t)\,dt+\sqrt{\nu_k({\theta}_{k},t;\boldsymbol{v}_{k-1}(\boldsymbol{\theta}_{k-1}))}\,dW_t\\
	{X}_{T_0}(\boldsymbol{\theta})&=&x_0
\end{array}	\label{SDE_mx41}
\end{eqnarray}
for $t\in[T_{k-1},T_k)$ and $k=1,\ldots,n$. 

Next we outline a parametrization such that the resulting \mx MGD has the same probability as \eqref{falvol2}. We consider \mx MGP $\hat{\mathcal{M}}$ based on the series of parametrization $\{\mathcal{M}_i\}_{i=k}^n$ and the starting value $\tilde{X}_{T_{k-1}}$, i.e., $\hat{\mathcal{M}}=(\hat m, \hat\nu(\hat{\boldsymbol{\theta}},t),T_{k-1},\tilde{X}_{T_{k-1}},\Theta_k)$ where the mixing function is $\hat m(\hat{\boldsymbol{\theta}})=\prod_{i=k}^n m_i\left(\theta_{i}\right)$ with $\hat{\boldsymbol{\theta}}_i=(\theta_k,\ldots,\theta_i)$, $\hat{\boldsymbol{\theta}}=\hat{\boldsymbol{\theta}}_n$; the parametrization is
$\hat\nu(\hat{\boldsymbol{\theta}},t)= \nu_i(\theta_{i},t;\boldsymbol{v}_{k-1}(\boldsymbol{V}_{k-1}),\hat{\boldsymbol{v}}_{i-1}(\hat{\boldsymbol{\theta}}_{i-1}))$ with  $$\hat{\boldsymbol{v}}_{i}(\hat{\boldsymbol{\theta}}_{i})=(v_{k}(\boldsymbol{V}_{k-1},\hat{\boldsymbol{\theta}}_k),\ldots,v_{i}(\boldsymbol{V}_{k-1},\hat{\boldsymbol{\theta}}_i)),$$
for every $t\in[T_{i-1},T_{i})$ and $i=k,\ldots,n$. Then we can define the desired \mx MGD based on this parametrization as $(\hat{X},\hat{\mathcal{M}},\hat{\boldsymbol{V}})$ where $\hat{\boldsymbol{V}}=(V_k,\ldots,V_n)$. Note that, the underlying diffusion process of $\hat X$ is the constrained stochastic process of $X_t(\cdot)$ with the starting value $\tilde{X}_{T_{k-1}}$ and initial total variance $\boldsymbol{v}_{k-1}(\boldsymbol{V}_{k-1})$. Therefore, the underlying diffusion process has the same law given by \eqref{falvol2}. Consequently, we have
\begin{eqnarray}
\mathbb{P}\left(\hat{X}\in A\right)
=\mathbb{P}(X(\boldsymbol{V})\in A|X_{T_{k-1}}(\boldsymbol{V})=\tilde{X}_{T_{k-1}},\boldsymbol{v}_{k-1}(\boldsymbol{V}_{k-1}))
=\mathbb{P}\left(\tilde{X}\in A\left|\mathcal{F}_{T_{k-1}^-}\right.\right).\label{falvol3}\nn
\end{eqnarray}

Adapting notation from \cite{GLASSERMAN2011}, we call the implied volatility conditioning on $\mathcal{F}_{T_{k-1}^-}$ the fully-conditional implied volatility. The following corollary is the direct result of above arguments.

\begin{cor}\label{corcondvol1}
The fully-conditional implied volatility of \mx MGD $(\tilde{X},{\mathcal{M}},\boldsymbol{V})$. with respect to $\mathcal{F}_{T_{k-1}^-}$ equals the spot implied volatility of $(\hat{X},\hat{\mathcal{M}},\hat{\boldsymbol{V}})$.
\end{cor}

\subsection{Parametrization with Joint Probability Distribution} \label{sec53}

In this section, we show how to parametrize the \mx mixture diffusion given the joint probability distribution of total variance. Because the information from one finite dimensional distribution is insufficient to determine the mixture diffusion, we always use some simple interpretation method to bridge the parametrization between the maturities. Under a chosen interpretation method, the mixture diffusion is then uniquely determined by the parametrization at the given maturities. For this reason, our definition of \mx MGP uses only time-independent mixing function. The following proposition shows there exist an ``unique" \mx mixture diffusion such that the joint probability distribution of total variance has the exact same distribution.
\begin{prop}\label{propx1}
Assume that $Y_1,\ldots,Y_n$ is a vector of random variables satisfying $v_0\le Y_{1}\le\cdots\le Y_n$. Define the conditional probability
\begin{eqnarray}
F_k(\sigma_k|\boldsymbol{\sigma}_{k-1}):=\mathbb{P}(Y_k-Y_{k-1}\le\sigma_k|Y_1=\sigma_1,\cdots,Y_{k-1}=\sigma_{k-1}).\label{mjd01}
\end{eqnarray}

Assume that there exists an equivalent MGP $\mathcal{M}_k$ such that
\begin{eqnarray}
\mathcal{M}_k=(m_k,\nu_k(\theta,t;\boldsymbol{\sigma}_{k-1}),T_{k-1},1,\Theta_k)\sim
(F_k^\prime(\cdot;\boldsymbol{\sigma}_{k-1}),\theta/(T_k-T_{k-1}),T_{k-1},1,[0,\infty))\nn
\end{eqnarray}
for all possible $0\le v_0\le \sigma_{1}\le\cdots\le \sigma_n$. Let ${\mathcal{M}}$ be the \mx MGP based on the series of parametrization $(\mathcal{M}_1,\ldots,\mathcal{M}_n)$ and $(\tilde{X},\mathcal{M},\boldsymbol{V})$ be the \mx MGD based on the parametrization ${\mathcal{M}}$. 

Then $(v_{1}(\boldsymbol{V}_{1}),\cdots,v_{n}(\boldsymbol{V}_{n}))$ of \mx MGD $\tilde{X}$ has the same joint probability distribution as that of $(Y_1,\ldots,Y_n)$.
\end{prop}

{\noindent\bf Proof of Proposition \ref{propx1}:} Denote $M_k(\cdot)$ as the cumulative distribution function of $m_k(\cdot)$. The equation \eqref{MIXEQU21} of the equivalent relation yields
\begin{eqnarray}
\int_{T_{k-1}}^{T_k}\nu_k(M_k^{-1}(x),s;\boldsymbol{\sigma}_{k-1})\,ds=F^{-1}_k(x;\boldsymbol{\sigma}_{k-1})\nn
\end{eqnarray}
where $x\in[0,1)$. Plugging in $\boldsymbol{\sigma}_{{k-1}}=\boldsymbol{v}_{k-1}(\boldsymbol{V}_{k-1})$ and $x=M_k(V_{k})$, it yields
\begin{eqnarray}
v_{k}(\boldsymbol{V}_{k})-v_{k-1}(\boldsymbol{V}_{k-1})=
F^{-1}_k(M_k(V_{k});\boldsymbol{v}_{k-1}(\boldsymbol{V}_{k-1}))\nn
\label{volIter1}.
\end{eqnarray}
Consequently, we have
\begin{eqnarray}
\mathbb{P}\left(v_{k}(\boldsymbol{V}_{k})-v_{k-1}(\boldsymbol{V}_{k-1})\le x\bigg|\boldsymbol{v}_{k-1}(\boldsymbol{V}_{k-1})\right)
%&=&\mathbb{P}\left(F^{-1}_k(M_k(V_{k});\boldsymbol{v}_{k-1}(\boldsymbol{V}_{k-1}))\le x\bigg|\boldsymbol{v}_{k-1}(\boldsymbol{V}_{k-1})\right)\nn\\
&=&\mathbb{P}\left(M_k(V_{k})\le F_k(x;\boldsymbol{v}_{k-1}\left(\boldsymbol{V}_{k-1})\right)\bigg|\boldsymbol{v}_{k-1}(\boldsymbol{V}_{k-1})\right)\nn\\
&=&F_k(x;\boldsymbol{v}_{k-1}(\boldsymbol{V}_{k-1})).\label{mjd02}
\end{eqnarray}
Comparing equation \eqref{mjd02} with the definition \eqref{mjd01}, it shows that $(v_{1}(\boldsymbol{V}_{1}),\cdots,v_{n}(\boldsymbol{V}_{n}))$ have the same joint probability density distribution as that of $(Y_1,\ldots,Y_n)$.\noindent$\blacksquare$

Next we apply Proposition \ref{propx1} to stochastic volatility models and show that certain \mx MGD can have the same joint probability distribution function as that of stochastic volatility models at the given maturities. Our result applies to the following general form of stochastic volatility models for asset price
\begin{eqnarray}
\begin{array}{rll}
\ds\frac{\,dS_t}{S_t}&=&r(t)\,dt + \sqrt{\nu_t}\,dW_t\\
S_0&=&x_0
\end{array}
\label{localSV1} \end{eqnarray}
where we assume that $\nu_t$ is a positive regular diffusion process and is independent to the standard Brownian Motion $W_t$.

\begin{cor}\label{lem02} Let $T_0=0$ and define the integral process of the total variance as $I(t):=\int_0^t\nu_s\,ds$. Assume that $(v_{1}(\boldsymbol{V}_{1}),\cdots,v_{n}(\boldsymbol{V}_{n}))$ of the \mx MGD $(\tilde{X},\mathcal{M},\boldsymbol{V})$ has the same joint probability distribution function as that of $(I(T_1),\ldots,I(T_n))$. 

Then $(X_{T_1},\ldots,X_{T_n})$ and $(S_{T_1},\ldots,S_{T_n})$ has the same joint probability distribution function.
\end{cor}

{\noindent\bf Proof of Proposition \ref{lem02}:} Note that, for both \mx MGD \eqref{SDE_mx40} and the stochastic volatility model \eqref{localSV1}, given the full path of volatility, the asset price follows log-normal distributions. It is also clear that for both models, conditioning on the asset price at $T_k$ and the total variance at $T_{k}$ and $T_{k+1}$, the conditional distribution of the asset price at $T_{k+1}$ is log-normal:
\begin{eqnarray}
&&\mathbb{P}\left(\log(X_{T_{k+1}}/F(T_{k+1}))\in\,dx_{k+1}\left|\log(X_{T_k}/F(T_k))=x_{k},v_{k+1}(\boldsymbol{V}_{k+1})=\sigma_{k+1},v_{k}(\boldsymbol{V}_k)=\sigma_k\right.\right)\nn\\
&=&\mathbb{P}\left(\log(S_{T_{k+1}}/F(T_{k+1}))\in\,dx_{k+1}\left|\log(S_{T_k}/F(T_k))=x_{k},I_{k+1}=\sigma_{k+1},I_k=\sigma_k\right.\right)\nn\\
&=&\phi(x_{k+1}-x_{k},\sigma_{k+1}-\sigma_k)\,dx_{k+1}\label{condist1}
\end{eqnarray}
where $\phi(x,\sigma):=\frac{1}{\sqrt{2\pi\sigma}}\exp\left(-\frac{(x+\sigma/2)^2}{2\sigma}\right).$ Then we obtain the joint probability distribution of the asset price at $T_1,\ldots, T_n$ by integrating the condition distribution \eqref{condist1} over the joint probability distribution of the total variance. 
\begin{eqnarray}
&&\mathbb{P}\left(\log(S_{T_{1}}/F(T_{1}))\in\,dx_{1},\cdots,\log(S_{T_n}/F(T_k))\in x_{n}\right)\nn\\
&=&\prod_{k=1}^n\,dx_{k} \int_0^\infty\cdots\int_0^\infty\mathbb{P}(I(T_1)\in\,d\sigma_1,\cdots,I(T_n)\in\,d\sigma_n)\prod_{k=1}^n\phi(x_{k}-x_{k-1},\sigma_{k}-\sigma_{k-1})\,d\sigma_k\nn\\
&=&\mathbb{P}\left(\log(\tilde{X}_{T_1}/F(T_{1}))\in\,dx_{1},\cdots,\log(\tilde{X}_{T_n}/F(T_k))\in x_{n}\right)\nn
\end{eqnarray}
This proves that $(S_{T_1},\ldots,S_{T_n})$ has the same joint probability distribution function as that of $(X_{T_1},\ldots,X_{T_n})$. \noindent$\blacksquare$

\subsection{Parametrization with Risk Neutral Distributions} \label{sec52}

In this section, we consider how to parametrization \mx MGD such that its marginal distribution matches the risk-neutral distributions implied by spot vanilla options as wells as those implied by forward start options. We only consider the risk-neutral distribution at fixed maturities $T_1,\ldots,T_n$ because the market trades only a finite number of options and we can ensure that $\{T_1,\ldots,T_n\}$ covers all maturities of interest. 

The following proposition shows that the marginal distribution of $\tilde X_t/\tilde X_s$ is completely determined by the distribution of total variance in the period $[s,t]$. 

\begin{prop}\label{cor3a} Let $v_{s,t}=\int_{s}^t\nu(\boldsymbol{V},x)\,dx$ and $F(s,t)=\exp(\int_{s}^tr(x)\,dx)$. Denote the marginal distribution of $v_{s,t}$ as $f_{s,t}(\cdot):[0,\infty)\mapsto[0,\infty)$. 

Then the marginal distribution function of $\tilde{X}_t/\tilde{X}_s$ is
\begin{eqnarray}
\mathbb{P}\left(\tilde{X}_t/\tilde{X}_s\in\,dx\right)=\int_0^\infty \frac{f_{s,t}(\theta)}{\sqrt{2\pi\theta}x}\exp\left(\frac{(\log(x/F(s,t))-\theta/2)^2}{2\theta}\right)\,d\theta\,dx\label{fouriersolution02}.
\end{eqnarray}
\end{prop}

{\noindent\bf Proof of Proposition \ref{cor3a}:} Direct calculation shows
\begin{eqnarray}	
\mathbb{P}\left(\log(\tilde{X}_t/\tilde{X}_s)\in\,dx\right)
&=&\mathbb{E}\left(\mathbb{P}\left(\left.\log(\tilde{X}_t/\tilde{X}_s)\in\,dx\right|\boldsymbol{V}\right)\right)\nn\\
&=&\mathbb{E}\left(\mathbb{P}\left(\log(X_t(\boldsymbol{V})/X_s(\boldsymbol{V}))\in\,dx\right)\right)\nn\\
&=&\mathbb{E}\left(\frac{1}{\sqrt{2\pi v_{s,t}}x}\exp\left(\frac{(\log(x/F(s,t))-v_{s,t}/2)^2}{2v_{s,t}}\right)\right)\,dx\nn\\
&=&\int_0^\infty \frac{f_{s,t}(\theta)}{\sqrt{2\pi\theta}x}\exp\left(\frac{(\log(x/F(s,t))-\theta/2)^2}{2\theta}\right)\,d\theta\,dx\nn
\end{eqnarray} 
\noindent$\blacksquare$

First we show how to match the risk-neutral distributions implied by vanilla options. Denote $D_k(x)$ as the risk-neutral distribution of the asset price at $T_k$; and the marginal density function of $v_{k}(\boldsymbol{V}_{k})$ as $l_k(\cdot)$. Below we solve the marginal distribution $l_k(\cdot)$ from the risk-neutral distribution $D_k(x)$ by applying Proposition \ref{cor3a} with $s=0$ and $t=T_k$. Note that the total variance at time $T_k$ is $v_{k}(\boldsymbol{V}_{k})$, we have
\begin{eqnarray}
D_k(x)=\int_0^\infty \frac{l_k(\theta)}{\sqrt{2\pi\theta}x}\exp\left(\frac{(\log(x/F(T_k))-\theta/2)^2}{2\theta}\right)\,d\theta\nn
\end{eqnarray}
As in \eqref{fourierequ1}, we can similarly apply Fourier transform to the logarithm of the risk-neutral function and solve $l_k(\cdot)$ as
\begin{eqnarray}
l_k(x)=\mathcal{L}^{-1}(G_k)(x).\label{fouriersolution2}
\end{eqnarray} 
where $\ds G_k(\eta)=\mathcal{F}(E_k)\left(\sqrt{2\eta-1/4}-{i}/2\right)$ and $E_k(x)=F(T_{k})e^xD_k(F(T_k)e^x).$

Next we show how to match the risk-neutral distributions implied by forward start options. Particularly, we consider the risk-neutral distribution of $\tilde{X}_{T_k}/\tilde{X}_{T_{k-1}}$ derived from Cliquet options. Denote the risk-neutral distribution of $\tilde{X}_{T_k}/\tilde{X}_{T_{k-1}}$ as $\hat D_k(\cdot)$ and the marginal density function of $v_{k}(\boldsymbol{V}_{k})-v_{k-1}(\boldsymbol{V}_{k-1})$ as $\hat l_k(\cdot)$. In Proposition \ref{cor3a}, we let $s=T_{k-1}$ and $t=T_k$. Note that the total variance of $\tilde{X}_{T_k}/\tilde{X}_{T_{k-1}}$ is $v_{k}(\boldsymbol{V}_{k})-v_{k-1}(\boldsymbol{V}_{k-1})$. Similar to \eqref{fouriersolution2}, we obtain
\begin{eqnarray}
\hat l_k(x)=\mathcal{L}^{-1}(\hat G_k)(x).\label{fouriersolution3}
\end{eqnarray} 
where $\hat G_k(\eta)=\mathcal{F}(\hat E_k)\left(\sqrt{2\eta-1/4}-{i}/2\right)$ and $\hat E_k(x):=F(T_{k-1},T_{k})e^x\hat D_k(F(T_{k-1},T_k)e^x).$

Equations \eqref{fouriersolution2} and \eqref{fouriersolution3} gives the explicit formula of the marginal distributions of the total variance that is consistent with the market implied risk-neutral distributions. However, the information of marginal distributions along are insufficient for us to derive the parametrization for \mx MGD. Base on Proposition \ref{propx1}, we also need the joint probability distribution of the total variances that is consistent with those marginal distributions.  This is related to the well-known problem attributed to A. N. Kolmogorov in \cite{Makarov1981}: find the joint distribution of random variables $X$ and $Y$ such that the marginal distributions of $X$, $Y$ and $X+Y$ equal the given probability distributions. The existence of such joint probability distribution function 
has been proved in \cite{Makarov1981}, \cite{Ruschendorf1982} and \cite{Frank1987}. This problem is also a special case of finding Copula distributions with fixed marginal distributions and there is a wide range of Copula distributions to model this join probability distributions \cite{Roger2006}.

Therefore, we can assume there exists a join distribution for $(v_{k-1}(\boldsymbol{V}_{k-1}), v_{k}(\boldsymbol{V}_{k}))$ such that the marginal distribution functions of $v_{k-1}(\boldsymbol{V}_{k-1})$, $v_{k}(\boldsymbol{V}_{k})$ and $v_{k}(\boldsymbol{V}_{k})-v_{k-1}(\boldsymbol{V}_{k-1})$ are exactly $l_{k-1}(\cdot)$, $l_k(\cdot)$ and $\hat l_k(\cdot)$, respectively. Denote the resulting joint probability distribution as $f_k(x,y)$. If we further assume 
the sequence $v_{1}(\boldsymbol{V}_{1}),\cdots,v_{n}(\boldsymbol{V}_{n})$ is markovian, we obtain the desired joint probability distribution of $(v_{1}(\boldsymbol{V}_{1}),\cdots,v_{n}(\boldsymbol{V}_{n}))$ as 
\begin{eqnarray}
\mathbb{P}\left(v_{1}(\boldsymbol{V}_{1})\in\,dx_1,\cdots,v_{n}(\boldsymbol{V}_{n})\in\,dx_n\right)=l_1(x_1)\,dx_1\prod_{k=2}^n f_k(x_{k-1},x_k)/l_{k-1}(x_{k-1})\,dx_k.
\label{jointdist2}
\end{eqnarray}
We can then apply Proposition \ref{propx1} to the join probability distribution \eqref{jointdist2}. The marginal distributions of the resulting \mx MGD will match exactly with the risk-neutral distributions implied by both spot vanilla options and forward start options. That is the following theorem.

\begin{thm}\label{lem05} 
Assume that the joint probability distribution of $(v_{1}(\boldsymbol{V}_{1}),\cdots,v_{n}(\boldsymbol{V}_{n}))$ is defined in \eqref{jointdist2}; $\tilde{X}_t$ is the resulting \mx MGD from Proposition \ref{propx1}. Then $\tilde{X}_{T_k}$ has the marginal distribution function $D_k(x)$ and $\tilde{X}_{T_k}/\tilde{X}_{T_{k-1}}$ has the marginal distribution function $\hat D_k(\cdot)$ for $k=1,\ldots,n$.
\end{thm}

\section{Conclusion}

\mx Volatility Model has two advantages over the commonly used asset pricing models which allow it to fit exactly to both spot vanilla options and forward start options. First, it models the volatility with a free-functional form and the volatility function can be solved from the risk-neutral distribution. Any asset pricing models with finite dimensional parametrization can approximate the risk-neutral distributions, but not match exactly. With the free-functional parametrization, \mx Volatility Model yields the exact solution to the option market. This is not possible using the traditional asset pricing models with finite number of parameters. Secondly, \mx Volatility Model can model the joint distribution of variance at a given set of maturities so that it fits exactly to the Vol-of-Vol structure implied by the forward start options. For example, by adapting the distribution of the variance from suitable stochastic volatility models, \mx Volatility Model duplicates the forward dynamics of the stochastic process at the given maturities, and price derivatives identically as the stochastic volatility models if values of derivatives depend only on these maturities.

\section{Appendix}

{\noindent\bf Proof of Theorem \ref{thmR3}:} Our proof runs in parallel to that in Theorem 1.1 of \cite{Friedman1975}. However, we make many adjustments to accommodate the unbounded parametrization.

First we prove the existence of the solution. Let $N$ and $\Delta$ be positive numbers. Define 
\begin{eqnarray}
\phi_N(V)=\left\{
\begin{array}{ll}
1,&(N-1)\Delta\le f(V)< N\Delta\\
0,&\text{Otherwise}.
\end{array}
\right.\nn
\end{eqnarray} 
We define $Y_t^{(k,N)}$ by iteration as follows:
\begin{eqnarray}
\begin{array}{ll}
Y_t^{(0,N)}=x_0,&\\
Y_t^{(k+1,N)}=\ds x_0+\int_0^t\phi_N(V)\mu(Y_s^{(k,N)},s;V)\,ds+\int_0^t\phi_N(V)\sigma(Y_s^{(k,N)},s;V)\,dW_s.&
\end{array} 
\label{AppLemR1E0}
\end{eqnarray} 
We have the following lemmas regarding the upper bounds of $\left|Y_t^{(k+1,N)}-Y_t^{(k,N)}\right|^2$
\begin{lem}\label{AppLemR1} Define $Y_t^{(k,N)}$ as in \eqref{AppLemR1E0}, then
\begin{eqnarray}
\mathbb{E}\left|Y_t^{(k+1,N)}-Y_t^{(k,N)}\right|^2 \le \frac{(M_0t(N\Delta)^2)^{k+1}}{(k+1)!}(1+|x_0|)^2\mathbb{E}\left(\phi_N(V)\right)\label{AppLemR1E1}
\end{eqnarray}
where $M_0=2\tau_0+2$.
\end{lem}
\begin{lem}\label{AppLemR2} Define $Y_t^{(k,N)}$ as in \eqref{AppLemR1E0}, then
\begin{eqnarray}
\mathbb{E}\sup_{0\le t\le \tau_0}\left|Y_t^{(k+1,N)}-Y_t^{(k,N)}\right|^2 
&\le& (2\tau_0+8)(N\Delta)^2\int_0^{\tau_0}\mathbb{E}\left|Y_s^{(k,N)}-Y_s^{(k-1,N)}\right|^2\,ds\label{AppLemR1E11}
\end{eqnarray}
\end{lem}
Combining \eqref{AppLemR1E1} and \eqref{AppLemR1E11}, we have
\begin{eqnarray}
\mathbb{E}\sup_{0\le t\le \tau_0}\left|Y_t^{(k+1,N)}-Y_t^{(k,N)}\right|^2 
%&\le& (2\tau_0^2+8\tau_0)(N\Delta)^2\frac{(M\tau_0(N\Delta)^2)^{k}}{k!}(1+|x_0|)^2\mathbb{E}\left(\phi_N(V)\right)\nn\\
&\le&H \frac{(M(N\Delta)^2)^{k+1}}{k!} \mathbb{E}\left(\phi_N(V)\right)\nn
\end{eqnarray}
where $H=4\tau_0(1+|x_0|)^2$ and $M=2\tau_0(\tau_0+1)$.

Now we define $X_s^{(k)}$ iteratively as
\begin{eqnarray}
\begin{array}{l}
X_t^{(0)}=x_0\\
X_t^{(k+1)}=\ds x_0+\int_0^t\mu(X_s^{(k)},s;V)\,ds+\int_0^t\sigma(X_s^{(k)},s;V)\,dW_s
\end{array}\label{equlimit1}
\end{eqnarray} 
It is clear that $X_t^{(k)}=Y_t^{(k,N)}$, if $(N-1)\Delta\le f(V)< N\Delta$. Then we have
\begin{eqnarray}
\mathbb{E}\sup_{0\le t\le \tau_0}\left|X_t^{(k+1)}-X_t^{(k)}\right|^2
&\le&\sum_{N=1}^\infty \mathbb{E}\left[\sup_{0\le t\le \tau_0}\left|Y_t^{(k+1,N)}-Y_t^{(k,N)}\right|^2\right] \nn\\
&\le&H\sum_{N=1}^\infty\int_{(N-1)\Delta\le f(\theta)<N\Delta}\frac{(M(N\Delta)^2)^{k+1}}{k!}m(d\theta)\nn\\
&\le&H\int_\Theta \frac{[M(f(\theta)+\Delta)^2]^{k+1}}{k!}m(d\theta)\nn
\end{eqnarray}
Note that $\Delta$ is arbitrary. Let $\Delta\to0$, by Dominant Convergence Theorem, we have,
\begin{eqnarray}
\mathbb{E}\sup_{0\le t\le \tau_0}\left|X_t^{(k+1)}-X_t^{(k)}\right|^2
&\le&H\int_\Theta \frac{[Mf^2(\theta)]^{k+1}}{k!}m(d\theta)\label{XTBOUND1}
\end{eqnarray}
Hence 
$$\mathbb{P}\left\{\mathbb{E}\sup_{0\le t\le \tau_0}\left|X_t^{(k+1)}-X_t^{(k)}\right|>\frac1{2^k}\right\}\le 2^{2k}H \int_\Theta \frac{[Mf^2(\theta)]^{k+1}}{k!}m(d\theta)$$
Because 
\begin{eqnarray}
\sum_{k=0}^\infty 2^{2k} \int_\Theta \frac{[Mf^2(\theta)]^{k+1}}{k!}m(d\theta)
&=& H\int_\Theta Mf^2(\theta) e^{4Mf^2(\theta)} m(d\theta)\nn\\
&\le& H\int_\Theta e^{5Mf^2(\theta)} m(d\theta) < \infty \nn
\end{eqnarray}
The Borel-Cantelli Lemma implies 
$$\mathbb{P}\left\{\mathbb{E}\sup_{0\le t\le \tau_0}\left|X_t^{(k+1)}-X_t^{(k)}\right|>\frac1{2^k}\ i.o. \ \right\}=0$$
Then it follows that $X_0+\sum_{i=1}^{k-1}X_t^{(i+1)}-X_t^{(i)}=X_t^{(k)}$ converge uniformly in $t\in[0,\tau_0]$. Denote the limit as $\tilde X_t$. Based on the standard arguments from Theorem 1.1 of \cite{Friedman1975}, we can show
\iffalse 
\begin{eqnarray}
\begin{array}{lll}
\mu(X_t^{(k)},t;V)\to\mu(\tilde X_t,t;V)\ \ \ &\text{uniformly\ in\ } &t\in[0,\tau_0]\\
\sigma(X_t^{(k)},t;V)\to\sigma(\tilde X_t,t;V)\ \ \ &\text{uniformly\ in\ } &t\in[0,\tau_0]\\
\end{array}\nn
\end{eqnarray}
%\begin{eqnarray}\int_0^T\left|\sigma(X_t^{(k)},t;V)-\sigma(\tilde X_t^{(k)},t;V)\right|^2\,ds\overset{P}{\rightarrow}0\nn\end{eqnarray}
\fi
$\tilde X_t$ satisfies the equation
\begin{eqnarray}
\tilde X_t=x_0+\int_0^t\mu(\tilde X_s,s;V)\,ds+\int_0^t\sigma(\tilde X_s,s;V)\,dW_s.\nn\end{eqnarray} 

Next we show $\tilde{X}_t$ is square integrable. From \eqref{XTBOUND1},
\begin{eqnarray}
\mathbb{E}\left|X_t^{(k+1)}\right|^2
&\le&|x_0|^2+\sum_{i=1}^k\mathbb{E}\sup_{0\le t\le \tau_0}\left|X_t^{(i+1)}-X_t^{(i)}\right|^2\nn\\
&\le&H\sum_{i=0}^k\int_\Theta \frac{[Mf^2(\theta)]^{i+1}}{i!}m(d\theta)\nn\\
&\le&H\int_\Theta Mf^2(\theta)e^{Mf^2(\theta)}m(d\theta)\nn
\end{eqnarray}
Let $k\to\infty$, from Fatou's lemma, we have 
\begin{eqnarray}
\mathbb{E}\left|\tilde{X}_t\right|^2<H\int_\Theta Mf^2(\theta)e^{Mf^2(\theta)}m(d\theta)<\infty.\label{XTBOUND2}
\end{eqnarray}
Therefore $\tilde{X}_t$ is a strong solution for the SDE. 

To prove the uniqueness of the solution, we assume $\tilde{X}_t^{(1)}$ and $\tilde{X}_t^{(2)}$ are two solutions. Let $\sigma(V)=1$, if $f(V)\le N$; 0, otherwise. Note that $0\le\sigma(V)f^2(V)\le \sigma(V)N^2$. Taking the expectation of $\sigma(V)\left|\tilde{X}_t^{(1)}-\tilde{X}_t^{(2)}\right|^2$, we have
\begin{eqnarray}
&&\mathbb{E}\left(\sigma(V)\left|\tilde{X}_t^{(1)}-\tilde{X}_t^{(2)}\right|^2\right)\nn\\
&\le&2\mathbb{E}\left(\sigma(V)\left|\int_0^t\left(\mu(\tilde{X}_s^{(1)},s;V)-\mu(\tilde{X}_s^{(2)},s;V)\right)\,ds\right|^2\right)\nn\\
&&+2\mathbb{E}\left(\sigma(V)\left|\int_0^t\left(\sigma(\tilde{X}_s^{(1)},s;V)-\sigma(\tilde{X}_s^{(2)},s;V)\right)\,dW_s\right|^2\right)\nn\\
&=&2\mathbb{E}\left|\int_0^t\sigma(V)\left(\mu(\tilde{X}_s^{(1)},s;V)-\mu(\tilde{X}_s^{(2)},s;V)\right)\,ds\right|^2
+2\int_0^t\mathbb{E}\left(\sigma(V)\left|\sigma(\tilde{X}_s^{(1)},s;V)-\sigma(\tilde{X}_s^{(2)},s;V)\right|^2\right)\,ds\nn\\
&\le&2t\int_0^t\mathbb{E}\left(\sigma(V)f^2(V)\left|\tilde{X}_s^{(1)}-\tilde{X}_s^{(2)}\right|^2\right)\,ds
+2\int_0^t\mathbb{E}\left(\sigma(V)f^2(V)\left|\tilde{X}_s^{(1)}-\tilde{X}_s^{(2)}\right|^2\right)\,ds\nn\\
&\le&2(t+1)N^2\int_0^t\mathbb{E}\left(\sigma(V)\left|\tilde{X}_s^{(1)}-\tilde{X}_s^{(2)}\right|^2\right)\,ds\nn
 \end{eqnarray}
Let $g(t)=\ds \mathbb{E}\left(\sigma(V)\left|\tilde{X}_t^{(1)}-\tilde{X}_t^{(2)}\right|^2\right)$, then $g(t)$ satisfies,
	$$0\le g(t)\le K\int_0^tg(s)\,ds, \tab\tab g(0)=0$$
where $K=2(\tau_0+1)N^2$ is a constant. Therefore $g(t)\equiv0$. Let $N\to\infty$, by Dominant Convergence Theorem, we prove $\mathbb{E}\left|\tilde{X}_t^{(1)}-\tilde{X}_t^{(2)}\right|^2=0$, hence the uniqueness of the solution. \noindent$\blacksquare$

{\noindent\bf Proof of Lemma \ref{AppLemR1}:} 
When $k=0$, 
\begin{eqnarray}
\left|Y_t^{(1,N)}-Y_t^{(0,N)}\right|^2 \le 
2\left|\int_0^t\phi_N(V)\mu(x_0,s;V)\,ds\right|^2+2\left|\int_0^t\phi_N(V)\sigma(x_0,s;V)\,dW_s\right|^2
\nn
\end{eqnarray}
Note that $0\le\phi_N(V)f(V)\le N\Delta$. Applying the linear growth condition, we have
\begin{eqnarray}
&&\mathbb{E}\left|Y_t^{(1,N)}-Y_t^{(0,N)}\right|^2 \nn\\
&\le& 2\mathbb{E}\left|\int_0^t\phi_N(V)\mu(x_0,s;V)\,ds\right|^2+2\int_0^t\mathbb{E}\Big|\phi_N(V)\sigma(x_0,s;V)\Big|^2\,ds\nn\\
&\le& 2\mathbb{E}\left(\left|\int_0^t \phi_N(V)f(V)(1+|x_0|)\,ds\right|^2\right)+2\mathbb{E}\left(\int_0^t\phi_N(V)f(V)(1+|x_0|)^2\right)\,ds\nn\\
&\le& 2\mathbb{E}\left(\phi_N(V)\right)\left|\int_0^t (1+|x_0|)N\Delta\,ds\right|^2+2\mathbb{E}\left(\phi_N(V)\right)\int_0^t\Big((1+|x_0|)N\Delta\Big)^2\,ds\nn\\
&=& \mathbb{E}\left(\phi_N(V) \right)(1+|x_0|)^2 M_0t (N\Delta)^2\nn
\end{eqnarray}
where $M_0=2\tau_0+2$. 

Now assume \eqref{AppLemR1E1} holds true for $k=0,1,\ldots,m-1$. When $k=m$, note that
\begin{eqnarray}
\left|Y_t^{(m+1,N)}-Y_t^{(m,N)}\right|^2 &\le &
2\left|\int_0^t\phi_N(V)\left[\mu(Y_s^{(m,N)},s;V)-\mu(Y_s^{(m-1,N)},s;V)\right]\,ds\right|^2\nn\\
&+&2\left|\int_0^t\phi_N(V)\left[\sigma(Y_s^{(m,N)},s;V)-\sigma(Y_s^{(m-1,N)},s;V)\right]\,dW_s\right|^2
\tab\tab
\label{AppLemR1E12}\end{eqnarray}
From the Lipschitz condition, we have
\begin{eqnarray}
&&\mathbb{E}\left|Y_t^{(m+1,N)}-Y_t^{(m,N)}\right|^2 \nn\\
&\le& 2\mathbb{E}\left|\int_0^t\phi_N(V)f(V)\left|Y_s^{(m,N)}-Y_s^{(m-1,N)}\right|\,ds\right|^2\nn\\
&&+2\mathbb{E}\int_0^t\phi_N(V)\left|\sigma(Y_s^{(m,N)},s;V)-\sigma(Y_s^{(m-1,N)},s;V)\right|^2\,ds\nn\\
&\le& 2(N\Delta)^2t\mathbb{E}\int_0^t\left|Y_s^{(m,N)}-Y_s^{(m-1,N)}\right|^2\,ds+2(N\Delta)^2\mathbb{E}\int_0^t\left|Y_s^{(m,N)}-Y_s^{(m-1,N)}\right|^2\,ds\nn\\
&\le& M_0(N\Delta)^2\int_0^t\mathbb{E}\left|Y_s^{(m,N)}-Y_s^{(m-1,N)}\right|\,ds\nn
\end{eqnarray}
Now plugging \eqref{AppLemR1E1} into the right side of the inequality, we have
\begin{eqnarray}
\mathbb{E}\left|Y_t^{(m+1,N)}-Y_t^{(m,N)}\right|^2
&\le& M_0(N\Delta)^2\int_0^t \frac{(M_0s(N\Delta)^2)^{m}}{(m)!}(1+|x_0|)^2\mathbb{E}\left(\phi_N(V)\right)\,ds\nn\\
&=&\frac{(M_0t(N\Delta)^2)^{m+1}}{(m+1)!}(1+|x_0|)^2\mathbb{E}\left(\phi_N(V)\right)\nn
\end{eqnarray}
By induction, we complete our proof. \noindent$\blacksquare$

{\noindent\bf Proof of Lemma \ref{AppLemR2}:} From \eqref{AppLemR1E12}, we have 
\begin{eqnarray}
\sup_{0\le t\le \tau_0}\left|Y_t^{(m+1,N)}-Y_t^{(m,N)}\right|^2 &\le &
2\tau_0(N\Delta)^2\int_0^{\tau_0}\left|Y_s^{(m,N)}-Y_s^{(m-1,N)}\right|^2\,ds\nn\\
&&+2\sup_{0\le t\le \tau_0}\left|\int_0^t\phi_N(V)\left[\sigma(Y_s^{(m,N)},s;V)-\sigma(Y_s^{(m-1,N)},s;V)\right]\,dW_s\right|^2\nn
\end{eqnarray}
Theorem 4.3.6 of \cite{Friedman1975} shows 
\begin{eqnarray}
&&\mathbb{E}\left\{\sup_{0\le t\le \tau_0}\left|\int_0^t\phi_N(V)\left[\sigma(Y_s^{(m,N)},s;V)-\sigma(Y_s^{(m-1,N)},s;V)\right]\,dW_s\right|^2\right\}\nn\\
&\le& 4 \mathbb{E}\int_0^t\phi_N(V)\left|\sigma(Y_s^{(m,N)},s;V)-\sigma(Y_s^{(m-1,N)},s;V)\right|^2\,dt\nn\\
&\le& 4 (N\Delta)^2\int_0^{\tau_0}\mathbb{E}\left|Y_s^{(m,N)}-Y_s^{(m-1,N)}\right|^2\,ds\nn
\end{eqnarray}
Therefore
\begin{eqnarray}
\mathbb{E}\sup_{0\le t\le \tau_0}\left|Y_t^{(m+1,N)}-Y_t^{(m,N)}\right|^2 
&\le& (2\tau_0+8)(N\Delta)^2\int_0^{\tau_0}\mathbb{E}\left|Y_s^{(m,N)}-Y_s^{(m-1,N)}\right|^2\,ds\nn
\end{eqnarray} 
\noindent$\blacksquare$

\end{document}